\documentclass[twocolumn,superscriptaddress,showpacs,nofootinbib,notitlepage,preprintnumbers,secnumarabic,amssymb, nobibnotes, aps, prl]{revtex4-2}
\usepackage[utf8]{inputenc}
\usepackage{graphicx}
\usepackage{latexsym,amsmath,amssymb,amsthm,lmodern,float,url,bm}
\usepackage{natbib}
\usepackage{color}
\usepackage{microtype}
\usepackage{import}
\usepackage{bbold}
\usepackage[plain]{fancyref}
\usepackage{varioref}
\usepackage{slashed}
\usepackage{multirow}
\usepackage{tikz}
\usepackage{braket}
\usetikzlibrary{shapes}
\usetikzlibrary{positioning}
\usepackage[normalem]{ulem}
\usepackage{bbm}
\usepackage{qcircuit}

\usepackage{graphics}
\usepackage{amsfonts}
\usepackage{multirow}
\usepackage{natbib}
\usepackage{ulem}
\usepackage[colorlinks=true,backref=false, linktocpage=true,
citecolor=blue,urlcolor=blue,linkcolor=blue,pdfpagemode=UseOutlines]{hyperref}
\hypersetup{%
  bookmarksnumbered=true,
  pdftitle = {},
  pdfsubject = {},
  pdfauthor = {},
  pdfkeywords = {}
}

\begin{document}
\preprint{FERMILAB-PUB-23-018-SQMS-T}
\title{Robustness of Gauge Digitization to Quantum Noise}
\author{Erik J. Gustafson}
\email{egustafs@fnal.gov}
\affiliation{Fermi National Accelerator Laboratory, Batavia,  Illinois, 60510, USA}
\author{Henry Lamm}
\email{hlamm@fnal.gov}
\affiliation{Fermi National Accelerator Laboratory, Batavia,  Illinois, 60510, USA}
\date{\today}
\begin{abstract}
Quantum noise limits the use of quantum memory in high energy physics simulations. In particular, it breaks the gauge symmetry of stored quantum states. We examine this effect for abelian and nonabelian theories and demonstrate that optimizing the digitization of gauge theories to quantum memory to account for noise channels can extend the lifetime before complete loss of gauge symmetry by $2-10\times$ over some other digitizations. These constructions also allow for quantum error correction to integrate the symmetries of quantum fields and prioritize the largest gauge violations.
\end{abstract}
\maketitle

Quantum computers can tackle high energy physics (HEP) problems beyond the reach of classical methods if noise can be controlled~\cite{Feynman,Jordan_2018,Jordan:2011ci,cohen2021quantum,Davoudi:2022uzo}. In particular the symmetry breaking by noise can spoil the predictive power of simulations by changing universality classes~\cite{Singh:2019jog,Singh:2019uwd,Bhattacharya:2020gpm,Zhou:2021qpm,2022PhRvD.105k4508A,Caspar:2022llo,Alexandru:2022son}.  Thus sufficiently preserving these symmetries is crucial~\cite{PhysRevA.99.042301,Raychowdhury:2018osk,2021PRXQ....2d0311H,2020arXiv200512688L,2022arXiv220607444M,2021arXiv211008041V,Halimeh:2022mct}; both in the current noisy era and the error-corrected future. An important example is gauge symmetry - invariance under local continuous transformations-- which for the Standard Model is $U(1)\times SU(2)\times SU(3)$.

Fault-tolerant quantum computers are many years off, and their form remains undecided. Different layouts and mixed architectures (combining qubits and qudits)
are being investigated~\cite{2022arXiv220300634A,B_kkegaard_2019,Yurtalan_2020,2022NatCo..13.3994B,2018OptL...43.5765Y,ciavarella2021trailhead,Gokhale_2019,baker2020efficient,https://doi.org/10.48550/arxiv.2208.04745,Alam:2022crs}. Further questions persist around the merits of quantum memory such as qRAM~\cite{PhysRevLett.100.160501,PhysRevA.78.052310,PhysRevA.86.010306,2015NJPh...17l3010A,PhysRevA.102.032608,2021PRXQ....2b0311H, 2022arXiv220603505C} and efficient quantum error correction (QEC)~\cite{Bravyi:1998sy,Breuckmann:2021yvk,Tan:2022oax,Skoric:2022bvc} where between $\mathcal{O}(10^{1-5})$ physical qubits per logical qubit appear necessary for fault tolerance~\cite{google_2020,ibm_2022,IONQPost,Girvin:2021txg}. Meanwhile, quantum advantage for HEP problems may be possible by designing them to be robust to certain errors, using partial QEC~\cite{2008PhRvA..78a2337S,https://doi.org/10.48550/arxiv.2205.13454,2021arXiv211102345C,10.1109ISCA45697.2020.00053,PhysRevLett.77.198,2016arXiv161006169F,2022arXiv220304948H}, hardware-aware embeddings~\cite{2022arXiv220510596L,2022arXiv220406369B,2020arXiv201003397N}, or biased-noise qubits~\cite{PhysRevA.78.052331,PhysRevLett.120.050505}.  Further robustness may come from mixed architectures. We show that lattice gauge theory (LGT) simulations will benefit from all these tools. 

Gauge bosons live in infinite-dimensional Hilbert spaces. Many digitization proposals have been studied (see Sec VI.b of~\cite{Bauer:2022hpo}); however a consideration of the behavior of physical hardware is missing. In particular, quantum noise can be connected to gauge-violating operators~\cite{Halimeh:2019svu,Halimeh:2020kyu,Halimeh:2020dft,PhysRevLett.125.240405,VanDamme:2020rur,Halimeh:2020xfd,Bonati:2021vvs,Bonati:2021hzo}; this insight can determine the robustness of a digitization to noise and extend the lifetime of gauge states in memory.

In this letter, we present digitizations of gauge theories onto quantum memory which are more robust to gauge-violating noise. This is achieved by encoding the group structure onto individual registers, and then embedding the most susceptible to symmetry breaking onto the cleanest qubits for current hardware. 
While this analysis is generally applicable, for concreteness we use the discrete subgroup approximation~\cite{Alexandru_2019,Ji_2020,2021arXiv210813305S,2022PhRvD.105k4508A,2022arXiv220315541G,2022arXiv220812309G}. 
With this truncation, freedom remains in encoding and embedding the group elements onto the quantum memory.
One can encode the elements onto qubits as binary integers~\cite{shaw_2020, Muschik_2017, Kaplan_2020,2021arXiv211108015B} or Gray codes~\cite{2004PhRvL..92q7902V,2020npjQI...6...49S,2021PhRvA.103d2405D,2021arXiv210308056C}.  
This analysis also provides LGT-specific stabilizers for QEC. We compare our digitization versus other methods with analytic results and simulations. For both abelian and nonabelian theories, our optimized digitization performs better than other methods to a degree similar to performing partial QEC.

Quantum noise can be described by a map on the system's density matrix, $\rho$.  This noise can be broadly classified as coherent or decoherent; we consider a subset of decoherent noise known as stochastic noise onto which any noise channel can algorithmically be transformed \cite{PhysRevX.7.021050,2013PhRvA..88a2314G,2018efficienttwirling,2016efficienttwirling,katabarwa2017dynamical,gutierrez2016errors,ouyang2011channel,ghosh2012surface,silva2008scalable}. Such error channels have multiple noise operators with independent probabilities $p_{i}$. We denote the application of these noise operators, $\hat{\mathcal{N}}_i$, on $\rho$ by the functional $\mathcal{F}(\rho, \hat{\mathcal{N}}_i) \equiv \hat{\mathcal{N}}_i \rho \hat{\mathcal{N}}_i^{\dagger}$.  The map of a noise channel on a register storing $\rho$ is given by  
\begin{equation}
    \label{eq:depolqubit}
    \mathcal{E}_{\mathcal{N}_i}(\rho) = \left(1 - \sum_{\hat{\mathcal{N}}_i} p_i\right) \rho + \sum_{\hat{\mathcal{N}}_i} p_i\,\mathcal{F}(\rho, \hat{\mathcal{N}}_i).
\end{equation}

Stochastic errors for qubits come in the form of bit, phase, and bit-and-phase flips denoted by the Pauli operators $\hat{X}_a$, $\hat{Z}_a$, and $\hat{Y}_a$; the subscripts indicate which qubit is acted on.  For later conciseness, we define $\hat{B}_a\in\{\hat{X}_a,\hat{Y}_a\}$.  
A \emph{depolarizing channel}, $\mathcal{E}_{\mathbf{1}}(\rho)$, is described by $\hat{\mathcal{N}}_i\in \lbrace \hat{X}_a, \hat{Y}_a, \hat{Z}_a|\,\forall\,a\rbrace$ and all $p_i$ are equal~\cite{nielsen_chuang_2010} . When only a phase flips can occur $\hat{\mathcal{N}}_i\in\lbrace\hat{Z}_a|\,\forall\,a\rbrace$ and it is called a \emph{dephasing channel}, $\mathcal{E}_{\mathbf{Z}}(\rho)$. 

For qudits with $V$ states, $\hat{X}_a$, $\hat{Y}_a$, and $\hat{Z}_a$ generalize to $\hat{X}^{(v,w)}_a$, $\hat{Y}^{(v,w)}_a$, $\hat{Z}^{(v)}_a$; the superscripts correspond to the qudit states that are acted on. Additionally, two new error channels exist:  \emph{clock shifts} and \emph{phase shifts}~\cite{2022arXiv220104546G, Gokhale_2019,2018PhRvA..98e2316M,2011PhRvA..84a2321K}. 
Clock shifts cyclically permute the states,
\begin{equation}
\hat{\chi}_a^{m} = \sum_{v=0}^{V - 1} |v\rangle \langle (v + m)\;\mathrm{mod}\; V|
\end{equation}
and phase shifts act as a conjugate operator to $\hat{\chi}^m_a$,
\begin{equation}
\hat{\mathcal{V}}_a^{m} = \sum_{v=0}^{v- 1} e^{2\pi i v m / V} |v\rangle\langle v|, 
\end{equation}
where $m$ are powers of $\hat{\mathcal{N}}_i$ i.e. $\hat{\chi}_a^2=\hat{\chi}_a\hat{\chi}_a$. The qudit depolarizing channel $\mathcal{E}_{\mathbf{1},q}(\rho)$ has $\hat{\mathcal{N}}_i \in \lbrace \hat{X}_a^{(v, w)}, \hat{Y}_a^{(v, w)}, Z^{(v)}_a, 
\hat{\chi}_a^m,\hat{\mathcal{V}}_a^m|\, \forall\,a,v<w<V,m<V\rbrace$ for a total of $3 V^2 - 5 V + 4$  operators.
The error channel excluding generalized Paulis $\mathcal{E}_{\chi\mathcal{V}}(\rho)$ has $V^2$  operators.
A final channel, $\mathcal{E}_{\mathcal{V}}(\rho)$, considers only the $V$ phase shifts. Having listed the error channels considered here, we move onto how they manifest in LGT simulations.

\begin{figure}
    \includegraphics[width=0.7\linewidth]{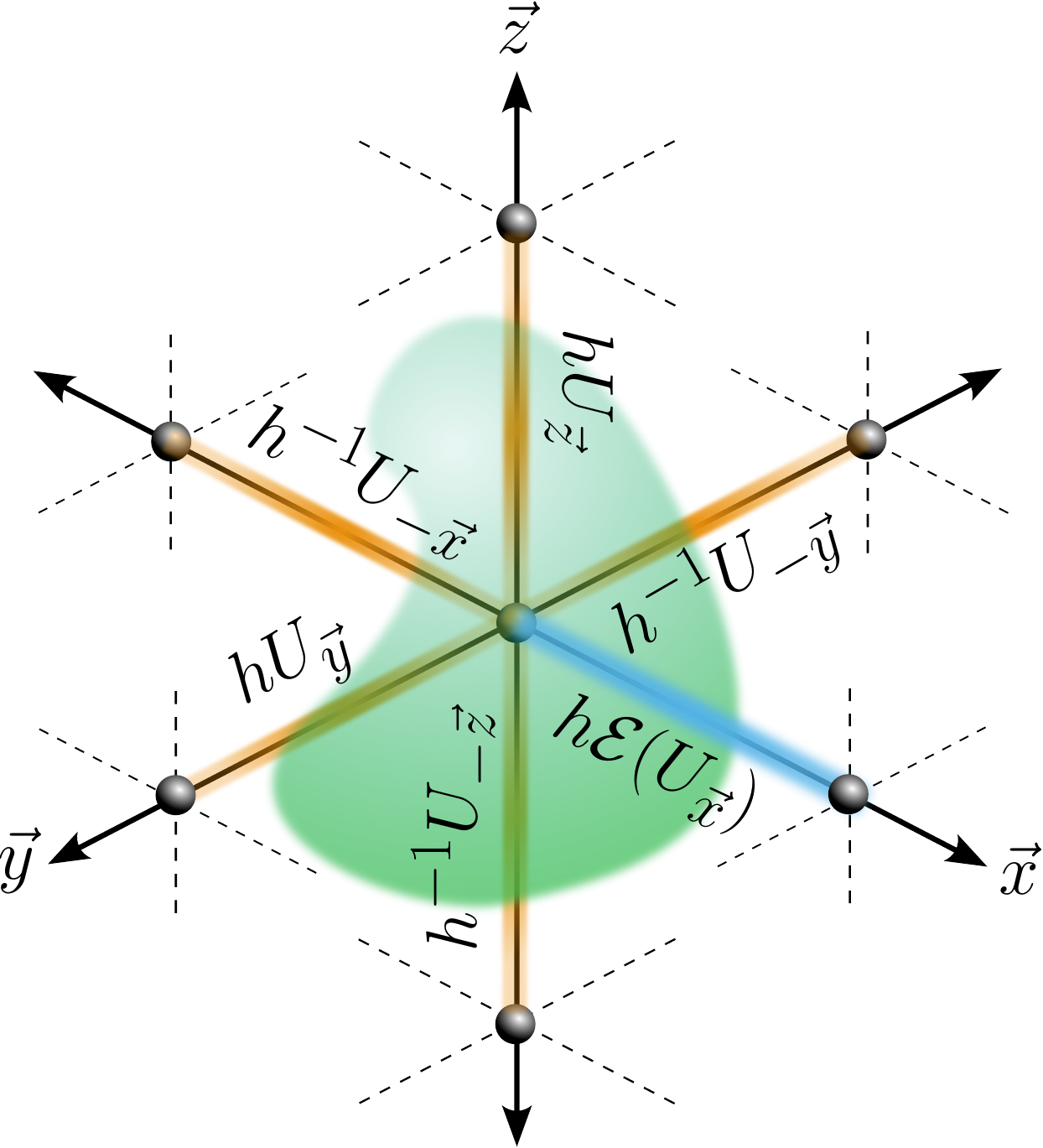}
\caption{Pictorial representation of Gauss's law, Eq.~(\ref{eq:gaugetransformationoperator}). $U_{\vec{x},i}\equiv U_{i}$ is the group element represented on a link, $\mathcal{E}(U_{\vec{x}})$ indicates the occurrence of noise on a link register.}
\label{fig:latticeintro}
\end{figure}

In Hamiltonian LGT, we discretize space. Excluding matter, the Hamiltonian is built from gauge links $U_{\vec{x},i}$ --- elements $h$ of the group $\mathbbm{H}$ which are integrals of the fields between sites $\vec{x}$ and $\vec{x}+\vec{i}$.  On a quantum computer, each link has a register $|U_{\vec{x},i}\rangle$ made of one or more qudits. 

The physical Hilbert space of states is the subset of all possible configurations $\prod_{\vec{x},i}a_{\vec{x},i}|U_{\vec{x},i}\rangle$ which obey Gauss's Law, i.e. gauge transformations around any site leave the state unchanged. The operator implementing gauge transformations by $h$ in $s$ spatial dimensions is
\begin{equation}
\label{eq:gaugetransformationoperator}
    \hat{G}(h)\prod_{i=1}^{s}|U_{\vec{x}, i}\rangle\otimes|U_{\vec{x} - \hat{i}, i}\rangle =
    \prod_{i=1}^{s}|h U_{\vec{x}, i}\rangle\otimes|U_{\vec{x} - \hat{i}, i} h^{\dagger}\rangle.
\end{equation}
The local operators $\hat{G}_i(h)$ construct Eq.~(\ref{eq:gaugetransformationoperator}) via $\hat{G}(h)=\prod_{i=1}^d\hat{G}_i(h)\otimes\hat{G}_{-i}(h^\dag)$ and necessarily commute with the Hamiltonian. We can determine how $\hat{\mathcal{N}}_i$ causes gauge violations for a digitization of $U_{\vec{x}, i}$ by checking each commutator  $[\hat{G}_i(h),\hat{\mathcal{N}}_i]$. The set of $h$ for which the commutator remains zero corresponds to the remnant symmetry $\mathbbm{G}\in\mathbbm{H}$ unbroken by the noise. A pictorial representation of gauge-violating noise is shown in Fig.~\ref{fig:latticeintro}.

For discrete groups, the group elements $h_d$ can be mapped to the integers $d=[0,|\mathbbm{H}|-1]$ where $|\mathbbm{H}|$ is the group dimension. One mapping uses ordered-products of group generators $\lbrace\lambda_k\rbrace$ raised to integer exponents $\{a_k\}$:
\begin{align}
\label{eq:groupelement}
h_{\lbrace o_k \rbrace} =& \prod_k \lambda^{o_k}_k = h_d
\end{align}
where $0\leq o_k<O_k$. $O_k$ is at most the generator's order $\lambda_k^{O_k}=\mathbbm{1}$ but lower when redundancies occur such as $\lambda_1^{O_1}=\lambda_2$.  An integer mapping is then defined as
\begin{align}
\label{eq:intrep}
 d=o_k + O_{k}(o_{k-1}+O_{k-1}(\hdots+O_2(o_1+O_1o_0))).
\end{align}
We consider two ways of encoding $d$ onto quantum memory.  One method decomposes $d$ via Eq.~(\ref{eq:intrep}) where $\{O_k\}$ are replaced by the dimensionality of the qudits $\{D_k\}$. This spans layouts including $\lceil\log_2(|{\mathbbm{H}|})\rceil$ qubits, a single $D_k=|\mathbbm{H}|$ qudit, or mixed memory where $\{D_k\}=\{O_K\}$. Another way, Gray codes, instead map $d$ to memory while minimizing bit flips. A $D=4$ example using qubits is $|0\rangle\rightarrow|00\rangle$, $|1\rangle\rightarrow|01\rangle$, $|2\rangle\rightarrow|11\rangle$, and $|3\rangle\rightarrow|10\rangle$. 

The cyclic groups, $\mathbbm{Z}_N$, naturally approximate $U(1)$.  $\mathbbm{Z}_N$ contains subgroups $\mathbbm{Z}_M$ where $M$ is each integer factor of $N$. Elements are generated via $h_d=\omega_N^d$ where $\omega_N$ is the $N^{th}$ root of unity. These $h_d$ can be encoded with $\lceil\log_2(N)\rceil$ qubits or a single $V=N$ qudit.
The $\hat{G}_i(h_d)$ for the binary encoding of $\mathbbm{Z}_N$ are
\begin{equation}
    \hat{G}_i(h_d) = \sum_{n = 0}^{N - 1} |(n + d) \text{ mod } N\rangle\langle n|
\end{equation} 
which can be mapped to quantum gates. As an example, for binary-encoded $\mathbbm{Z}_4$ on qubits:
\begin{align}
    \hat{G}_i(h_1) & = \frac{1}{2}\Big(\hat{X}_2 - i \hat{Y}_2 + \hat{X}_1(\hat{X}_2 + i \hat{Y}_2)\Big)\notag\\
    \hat{G}_i(h_2) &= \hat{X}_1,~\hat{G}_i(h_3) = \hat{G}_i(h_1)^{\dagger},~\hat{G}_i(h_0) = \mathbbm{1}_{4\times4}.
\end{align}
Of these $[\hat{G}_i(h_d), \hat{X}_1] = 0$; therefore, no symmetry is broken. Together $ [\hat{G}_i(h_2),\hat{Z}_2]=0$ and $[\hat{G}_i(h_1),\hat{Z}_2] \neq 0$ imply only a $\mathbbm{Z}_2$ symmetry is preserved. Since $\hat{Z}_1$ commutes only with $\hat{G}_i(h_0)$, it leaves no remnant symmetry.

Enumerating all commutators determines the symmetry breaking pattern for an encoding and leads to Table~\ref{tab:qubittable}. For a binary encoding of $\mathbbm{Z}_{2^n}$, $\hat{X}_a$ breaks the symmetry to $\mathbbm{Z}_{2^{n - a}}$, $\hat{Z}_a$ and $\hat{Y}_{a\neg 0}$ ($\neg$ being the negation sign) reduces the symmetry to $\mathbbm{Z}_{2^{n-1}}$, with $\hat{Y}_0$ leaving none. Most channels leave only a $\mathbb{Z}_2$ symmetry when Gray encoded. For $N \neq 2^n$ the noise breaking is generically worse due to ``forbidden" states -- states in the memory that represent no group element. For example, no remnant symmetries ever survive for either encoding when $N$ is odd. One $V=N$ qudit naturally encodes $h_n$ on $|n\rangle$. 
For this encoding $\hat{\mathcal{V}}^m$ leave a remnant $\mathbbm{Z}_2$ and $\hat{\chi}^m$ preserves the full $\mathbbm{Z}_N$. 

This symmetry breaking can be connected with the logical code states of LGT-specialized QEC. Since $\hat{G}_i(h)$ leave these states invariant, they form a stabilizer set. Using Table~\ref{tab:qubittable}, one can prioritize channels for partial QEC. This prioritization can greatly reduce resources since the number of physical qubits can scale from linearly to exponentially with logical qubits~\cite{Bravyi:1998sy,Breuckmann:2021yvk,Tan:2022oax,Skoric:2022bvc}. Phase transitions can separate theories with weak and strong gauge violations and different operators~\cite{Halimeh:2019svu,Halimeh:2020kyu,Halimeh:2020dft,PhysRevLett.125.240405,VanDamme:2020rur,Halimeh:2020xfd,Bonati:2021vvs,Bonati:2021hzo} so one can consider correcting only errors necessary to be in the same phase. For example, QEC that only repairs bit flips on $a=0$ to the binary encoding changes the scaling of full symmetry breaking from $p$ to $p^2$ since at least 2 errors are required. 

We measure the robustness of an encoding, $\epsilon$, by defining the average fraction of $\mathbbm{H}$ which remains invariant after one application of $\mathcal{N}_i$:
\begin{equation}
\label{eq:del}
    \delta_{\mathbbm{H}}^{\epsilon}=\frac{1}{|\mathbb{H}|}\sum_{\mathcal{N}_i} p_{i}|\mathbb{G}|_i.
\end{equation}
Equation~(\ref{eq:del}) depends upon $p_{i}$, and higher $\langle \delta_{\mathbbm{G}}\rangle_\epsilon$ would be possible with noise-aware embeddings. If we assume $p_{i}=p$, the binary encoding yields
\begin{align}
    \delta_{\mathbbm{Z}_{2^n}}^b&=\left[\frac{1}{3}+\frac{1}{2n}-\frac{2^{1-n}}{3n}\right]p
    \end{align}
    while Gray encoding yields
\begin{align}
        \delta_{\mathbbm{Z}_{2^n}}^g&=\left[\frac{1}{6n}+2^{1-n}-\frac{2^{2-n}}{3n}\right]p.
    \end{align}
    For the case of a single $V=N$ state qudit, we find 
\begin{align}    
        \delta_{\mathbbm{Z}_{N}}^d&=\left[\frac{2+N}{N(1+2N)}\right]p.
\end{align}
Setting $N=2^n$ we find $\delta_{\mathbbm{Z}_{2^n}}^b> \delta_{\mathbbm{Z}_{2^n}}^g>\delta_{\mathbbm{Z}_{2^n}}^d$ for $n>1$.  In fact, $\delta_{\mathbbm{Z}_{2^n}}^b\geq\frac{p}{3}$ while the others to zero as $n\rightarrow\infty$.



\begin{table}
    \caption{$\mathcal{N}_i$ vs. $\mathbbm{G}$ for $U(1)$ subgroups: $\mathbbm{Z}_N$ where $N=2^n$.}
    \label{tab:qubittable}

    \begin{tabular}{ccc|c}
    \hline\hline
     Binary & Gray &Qudits & $\mathbbm{G}$ \\ \hline
         $\hat{Y}_0$ & $\hat{Y}_0$ & $\hat{B}^{(i,j)}$, $\hat{Z}^{(i)}$ &--- \\
         --- &  $\hat{B}_{a\neg 0}$, $\hat{Z}_a$ & $\hat{\mathcal{V}}^m$ & $\mathbbm{Z}_2$ \\
                  $\hat{X}_{a\neg 0}$&--- & --- & $\mathbbm{Z}_{2^{n - a}}$\\
         $\hat{Z}_a,~\hat{Y}_{a\neg 0}$ &  $\hat{X}_0$ & --- & $\mathbbm{Z}_{2^{n - 1}}$  \\
         $\hat{X}_0$ & --- & $\hat{\chi}^m$ & $\mathbbm{Z}_{N}$\\
                  \hline\hline
    \end{tabular}   
\end{table}

As error rates increase, multiple errors could occur before detection and $\delta_{\mathbbm{G}}^\epsilon$ becomes insufficient. On real devices the errors vary across qubits and can combine to further break or restore symmetries.  To study these effects, Monte Carlo simulations were performed by evolving noisy registers $|U_n\rangle$ and measuring the probability $\mathcal{P}_{\mathbbm{G}}(t)$ that remnant symmetries persist. 

\begin{figure}
    \centering
    \includegraphics[width=0.95\linewidth]{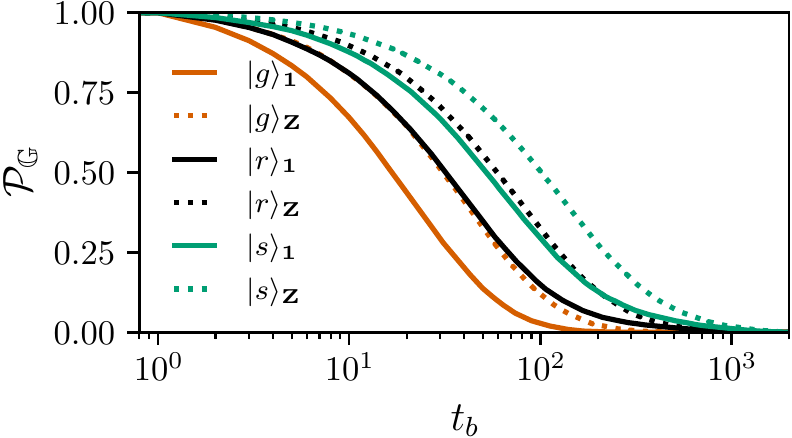}
    \caption{$\mathcal{P}_{\mathbb{G}}(t_b)$ for $\mathbbm{Z}_8$ versus $t_b$ using $|g\rangle$, $|r\rangle$, and $|s\rangle$ for depolarizing and dephasing channels.}
    \label{fig:z8qubits}
\end{figure}

We generated an ensemble of 250 $\mathbbm{Z}_8$ $|U_n\rangle$ for $\mathcal{E}_\mathbf{1}$ and $\mathcal{E}_\mathbf{Z}$ where the three qubits are assigned an independent set of $p_i$'s sampled from a truncated normal distribution with mean $1.2\%$ and standard deviation $0.5\%$, similar to rates on current transmon platforms~\cite{ibm_quantum}. Three embeddings for $\mathbbm{Z}_{8}$ are studied: an optimized binary embedding $|s\rangle$, a random binary embedding $|r\rangle$, and---as a benchmark---a random Gray code embedding $|g\rangle$. Optimized embeddings take the relative noisiness of the qubits within $|U_n\rangle$ into account when assigning qubits, while random embeddings are made without regard to noise. Since Gray encodings break nearly all symmetries, negligible improvement is gained by optimization. With the register ensembles, we build a time-dependent noise operator $\mathcal{A}(t)$.  Initialized to $\mathcal{A}(0)=\mathbbm{1}^{\otimes}$, we increment in discrete time $t$ and with probability $p_i$ multiply each $\hat{\mathcal{N}}_{i}$ onto $\mathcal{A}(t)$. Then all $[\hat{G}_i(h_d),\mathcal{A}(t)]$ are checked for remnant symmetries. When none remain, the simulation terminates and the symmetry-breaking time $t_b$ is recorded. We perform 100 simulations for each register. The resulting $\mathcal{P}_{\mathbbm{G}}(t_b)$ are shown in Fig.~\ref{fig:z8qubits}. In Tab.~\ref{tab:survivaltime}, we report $t_{1/2}$ defined as when $\mathcal{P}_{\mathbbm{G}}(t_{b})=1/2$. 

\begin{table}
\caption{$t_{1/2}$ for various groups, embeddings, and $\mathcal{E}_{\mathcal{N}_i}(\rho)$.}
\label{tab:survivaltime}
\begin{tabular}{ccccccc}
\hline\hline
$\mathbbm{H}$& $|g\rangle_1$ & $|g\rangle_{\mathbf{Z}}$ & $|r\rangle_1$ & $|r\rangle_{\mathbf{Z}}$ & $|s\rangle_1$ & $|s\rangle_{\mathbf{Z}}$\\\hline
$\mathbbm{Z}_8$ & 
20 & 32 & 33 & 76 & 80 & 105\\
$\mathbbm{BO}$ & 
--- & --- & 10 & 59 & 11 & 111\\
$\Sigma(36\times3)$ & 
--- & --- & 42 & 53 & 103 & 121\\
\hline\hline
\end{tabular}
\end{table}

The encoding via $|r\rangle_{\mathcal{E}}$ outperforms $|g\rangle_{\mathcal{E}}$, with even the noiser $|r\rangle_{\mathbf{1}}$ performing better than $|g\rangle_{\mathbf{Z}}$,  suggesting Gray encodings require more QEC to protect gauge symmetries. We find additional robustness is gained by even using the simple noise-awareness used to construct $|s\rangle_{\mathcal{E}}$. In both cases, cheap algorithmic changes can be as powerful as error correction in $\mathcal{E}_\mathbf{1}$. Together, the noise-aware binary digitization reduces the error-correcting resources required when storing quantum states.  Given the shallower circuits with Gray codes, simulations may benefit from running on computers with a QPU and memory using different encodings with the capability of in-transit swapping.

For subgroups of $SU(2)$, ordered-product encodings are known for the quaternions $\mathbb{Q}_8$, binary tetrahedral $\mathbb{BT}$, and binary octahedral $\mathbb{BO}$ group. These groups have $|\mathbbm{H}|=8, 24, 48$ respectively. Initial studies of quantum simulations exist for $\mathbb{Q}_8$~\cite{2021arXiv210813305S,2022arXiv220315541G} and $\mathbb{BT}$ \cite{2022arXiv220812309G}. These groups are generated by the quaternions $\mathbf{i} = iX,~\mathbf{j} = i Y,~\mathbf{k}=iZ$ and two others:
$\mathbf{t} = \frac{1}{\sqrt{2}}(\mathbbm{1} + \mathbf{i})$ and $\mathbf{l} = -\frac{1}{2}(\mathbbm{1} +\mathbf{i} + \mathbf{j} + \mathbf{k})$.
Since $\mathbf{i}^2=\mathbf{j}^2=\mathbf{k}^2=-\mathbbm{1}$ and $\mathbf{t}^2=\mathbf{i}$, each needs a single qubit with a shared sign qubit. In contrast, $\mathbf{l}^3=\mathbbm{1}$ and requires two qubits (with a forbidden state) or a qutrit.

\begin{table*}
    \caption{$\mathcal{N}_i$ vs. $\mathbbm{G}$ for three subgroups of $SU(2)$: $\mathbb{Q}_8$, $\mathbb{BT}$, and $\mathbb{BO}$.}
    \label{tab:SU2noiseTable}
    \begin{tabular}{cc|cc|c|c}
    \hline \hline
    \multicolumn{2}{c|}{$\mathbb{Q}_8$} & \multicolumn{2}{|c|}{$\mathbb{BT}$} & \multicolumn{1}{|c|}{$\mathbb{BO}$}&\\
   Qubits & Quoctit & Qubits & Quoctit + Qutrit & Qubits & $\mathbbm{G}$ \\ \hline
    
        $\hat{Z}_a$, $\hat{Y}_a$ & $\hat{\chi}_a^o$, $\hat{X}_a^{(v,\neg [v+1])}$, $\hat{\mathcal{V}}_a^m$, $\hat{Z}_a^{(v)}$, $\hat{Y}_a^{(v,w)}$ & $\hat{B}_{\{d_0,d_1\}}$& $\hat{B}^{(v,w)}_b$ , $\hat{Z}^{(0)}_b$,
    $\chi_b^o$, $\mathcal{V}_b^n$& $\hat{B}_{\{b_0,b_1\}}$, $\hat{Z}_a$, $\hat{Y}_a$ & ---\\
         $\hat{B}_{\{b,c\}}$&  $\hat{\chi}_a^e$, $\hat{X}_a^{(v, v+1)}$ & $\hat{B}_{\{b,c\}}$,$\hat{Z}_{\{b,c\}}$ & $\hat{\chi}_b^e$,$\hat{B}^{(n,m)}_a$  & $\hat{B}_{\{c,d,f\}}$&$\mathbbm{Z}_2$\\
         --- &--- & $Z_a$, $Y_a$ &  $\hat{\chi}_a^n$, $\hat{Z}^{(1)}_b$, $\hat{Z}^{(2)}_b$&--- & $\mathbbm{Z}_3$\\
         $\hat{Z}_{\{b,c\}}$&---&---&---&$\hat{Z}_{\{d,f\}}$& $\mathbbm{Z}_4$\\
         $\hat{X}_a$ &---& $\hat{Z}_{\{d_0,d_1\}}$ & $\hat{Z}^{(n)}_a$, $\hat{\mathcal{V}}_a^n$ & $\hat{Z}_{\{b_0,b_1\}}$ &  $\mathbb{Q}_8$\\
         ---& ---& $\hat{X}_a$ & --- & $\hat{Z}_c$ & $\mathbb{BT}$\\
         ---& ---& --- & --- & $\hat{X}_a$ & $\mathbb{BO}$ \\
         \hline\hline
    \end{tabular}
\end{table*}

$\mathbb{Q}_8$ can be stored in 3 qubits or a $V=8$ \emph{quoctit} via
\begin{equation}
    \label{eq:q8encodingqubits}
    h_d = (-1)^a \mathbf{i}^b \mathbf{j}^c\rightarrow |abc\rangle.
\end{equation} 
On the quoctit, we apply Eq.~(\ref{eq:intrep}) with alphabetical order. The effects of $\mathcal{E}_{\mathcal{N}}$ are enumerated in Table \ref{tab:SU2noiseTable}. For qubit registers, 7 of the 9 $\mathcal{N}_i$ leaves a remnant symmetry, with $\hat{X}_a$ leaving $\mathbbm{Q}_8$ unbroken.  Together this gives $\delta_{\mathbbm{Q}_8}^b=\frac{1}{3}$.  Compared to $\mathbbm{Z}_N$, symmetry breaking is larger for $Z_a$ compared to $X_a$. Thus the gauge group of interest can change the desired specialized hardware or error correction used. Alternatively, 68 of the 80 quoctit $\mathcal{N}_i$ break all symmetries.  The other 12 preserve only a $\mathbbm{Z}_2$ yielding $\delta_{\mathbbm{Q}_8}^d=\frac{3}{80}$; 9 times worse than the qubit encoding. 


$\mathbb{BT}=\mathbb{Q}_8 \rtimes \mathbbm{Z}_3$ can be encoded by extending the $\mathbb{Q}_8$ construction with $\mathbf{l}$:
\begin{equation}
    \label{eq:BTelements}
    h_d=(-1)^a \mathbf{i}^b \mathbf{j}^c \mathbf{l}^{d}.
\end{equation}
We consider two possible encodings: a five qubit register $|abcd_0d_1\rangle$ and a mixed register of a quoctit $|a\rangle$ for $\mathbb{Q}_8$ and a qutrit $|b\rangle$ for $\mathbf{l}$. For both registers, we observe channels can preserve nonabelian symmetries (Table~\ref{tab:SU2noiseTable}) with $\delta_{\mathbb{BT}}^b\approx0.16$ and $\delta_{\mathbb{BT}}^d\approx0.12$.

    

A final $SU(2)$ subgroup, $\mathbb{BO}$, requires 6 qubits with~\cite{inprepb0}
\begin{equation}
    \label{eq:qubitboencoding}
    h_d = (-1)^a \mathbf{l}^{b} \mathbf{t}^c \mathbf{j}^d \mathbf{k}^f \rightarrow |ab_0b_1cdf\rangle.
\end{equation}
For a qudit encoding, one could consider a qutrit for $\mathbf{l}$ with either two quoctits or one $V=16$ \emph{qudecasexit}.  We leave this investigation to future work as for both the generators can be ordered 24 ways and we anticipate strong dependence of symmetry on these choices. 

The symmetry breaking for the qubit encoding is provided in Table \ref{tab:SU2noiseTable}. 
Key features are that $\hat{X}_a$ break no symmetry, $Z_{b_0,b_1,c}$ leave remnant nonabelian groups, and $\delta_{\mathbb{BO}}^b=1/8$. Noisy $\mathbb{BO}$ registers were simulated with $t_{1/2}$ reported in Table~\ref{tab:survivaltime}. For $\mathcal{E}_\mathbf{1}$, noise optimization had negligible effect, but $t_{1/2}$ nearly doubled for $\mathcal{E}_\mathbf{Z}$.


The smallest crystal-like subgroup of $SU(3)$ with the center $\mathbbm{Z}_3$ is the 108-element $\Sigma(36\times3)$ discussed in~\cite{Grimus:2010ak}.  It has many subgroups including the dihedral group $\mathbbm{D}_3$, $\Delta(27)=(\mathbbm{Z}_3\times \mathbbm{Z}_3)\rtimes\mathbbm{Z}_2$, and $\Delta(54)=(\mathbbm{Z}_3\times \mathbbm{Z}_3)\rtimes\mathbb{S}_3$ where $\mathbb{S}_3$ is the symmetric group of three elements. The $\lambda_k$ used are $\omega_3$, $\mathbf{C}=\text{diag}(1,\omega_3,\omega_3^2)$,
\begin{align}
\label{eq:s108generators}
\textbf{E} &= \begin{pmatrix}
    0 & 1 & 0\\
    0 & 0 & 1\\
    1 & 0 & 0\\
    \end{pmatrix},\text{ and }\textbf{V} = \frac{1}{\sqrt{3}i}\begin{pmatrix}
    1 & 1 & 1\\
    1 & \omega_3 & \omega_3^2\\
    1 & \omega_3^2 & \omega_3\\
    \end{pmatrix}.
\end{align}

All generators have $O_k=3$ except $\mathbf{V}$ with $O_V=4$.  With these, an ordered-product encoding is:

\begin{equation}
    \label{eq:s108groupelements}
    h_d= \mathbf{\omega}_3^a \textbf{C}^b \textbf{E}^c \textbf{V}^d
\end{equation}

An eight-qubit encoding $|a_0a_1b_0b_1c_0c_1d_0d_1\rangle$ is possible where states with $|a\rangle,|b\rangle$, or $|c\rangle=|11\rangle$ are forbidden and error channels which mix into them breaking all symmetries. This is seen in Table~\ref{tab:s108noisetable}. $\Sigma(36\times3)$ can also be encoded onto 3 qutrits $|abc\rangle$ and a $V=4$ \emph{ququart} for $|d\rangle$ without forbidden states. $\hat{Z}_a$ break the symmetry of qubit encodings down to smaller nonabelian groups while most $\hat{B}_a$ leave no remnant symmetry. Every qudit channel preserves some symmetry, albeit mostly small abelian ones. Finally,  $\delta_{\Sigma(36\times3)}^b\approx0.15$ and $\delta_{\Sigma(36\times3)}^d\approx0.11$.

\begin{table}
    \centering
    \caption{$\mathcal{N}_i$ vs. $\mathbbm{G}$ for an $SU(3)$ subgroup: $\Sigma(36\times3)$.}
    \label{tab:s108noisetable}
    \begin{tabular}{ccc}
    \hline\hline
    Qubits & Qutrits+Ququart& $\mathbbm{G}$\\ \hline
         $\hat{B}_{\neg\{d_0,d_1\}}$& ---&---\\
         ---&$\hat{\mathcal{V}}_a^n$ & $\mathbbm{Z}_2$\\
         ---&$\hat{\chi}^n_c$, $\hat{\mathcal{V}}_{\{b,c\}}^n$, $\hat{B}^{(n, m)}_a$, $\hat{B}^{(0, m)}_{\{b,c\}}$& $\mathbbm{Z}_3$\\
         ---&$\hat{B}^{(m, \neg[m+2])}_d$,$\hat{Z}^{(n)}_a$,$\hat{Z}^{(\neg 0)}_{\{b,c\}}$ & $\mathbbm{Z}_3$\\
        ---&$\hat{Z}^{(0)}_{\{b,c\}}$, $\hat{B}^{(1, 2)}_{\{b,c\}}$, $\hat{B}^{(m, m+2)}_d$& $\mathbbm{Z}_6$\\ 
         $\hat{Z}_{\{a_0,a_1\}}$ &--- &$\mathbb{D}_3$\\
         ---&$\hat{\chi}^n_b$ & $\mathbbm{Z}_3 \times \mathbbm{Z}_3$\\
         ---&$\hat{\chi}^n_d$ & $\mathbbm{Z}_{12}$\\
         $\hat{Z}_{\{b_0,b_1,c_0,c_1\}}$ &---& $\mathbbm{Z}_3\rtimes\mathbbm{Z}_6$\\
         $\hat{Z}_{d_0}$, $\hat{B}_{d_0}$ & $\hat{\mathcal{V}}^n_d$, $\hat{Z}^{(n)}_d$ & $\Delta(27)$\\
         $\hat{Z}_{d_1}$, $\hat{Y}_{d_1}$ &---& $\Delta(54)$\\
         $X_{d_1}$ & $\hat{\chi}^n_a$ & $\Sigma(36\times3)$\\\hline\hline
    \end{tabular}
\end{table}

\begin{figure}
    \includegraphics[width=0.95\linewidth]{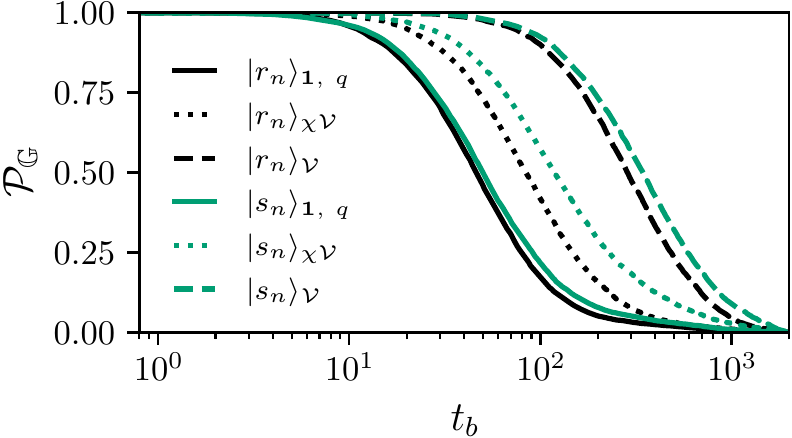}
    \caption{$\mathcal{P}_{\mathbbm{G}}$ for $\Sigma(36\times3)$ register vs. $t_b$ for qudit registers.\label{fig:quditss108}}
\end{figure}

We perform simulations for both encodings. For $|s_n\rangle_{\mathcal{E}}$ the qubits and qudits are ordered from noisiest to most quiet  $d,b,c,a$ with $q_1$ before $q_0$ for each qubit register following from Table \ref{tab:s108noisetable}. The $t_{1/2}$ for qubits are in Table~\ref{tab:survivaltime} and $\mathcal{P}_{\mathbbm{G}}(t_b)$ for qudits in Fig.~\ref{fig:quditss108}. Independent of noise model, we observe $|s_n\rangle_{\mathcal{E}}$ increase $t_{1/2}$ at least 2$\times$.

In this letter the robustness of gauge theory digitizations using discrete groups to quantum noise dependence was explored. Qubit and qudit encodings based on group structure and noise-aware embeddings can extend the range of quantum memory for storing states compared to methods which prioritize circuit depth. Partial error correction for lattice gauge theories with reduced resources are defined by prioritizing the largest gauge-violating error channels. These methods are applicable to other digitizations \cite{ciavarella2021trailhead,BROWER2004149,Beard_1998,2022arXiv220102412B} by identifying similar noise-symmetry relations. Future focus to matter fields should be made. Finally, threshold error rates for error correction using gauge-transformation stabilizers as in \cite{2021arXiv211205186R} should be determined for comparison with general QEC. Together, our results reduce the resources needed for quantum advantage in LFT simulations. 

\begin{acknowledgements}
The authors thank A.V. Grebe and R. Van de Water for helpful comments. This material is based on work supported by the U.S. Department of Energy, Office of Science, National Quantum Information Science Research Centers, Superconducting Quantum Materials and Systems Center (SQMS) under contract number DE-AC02-07CH11359. Fermilab is operated by Fermi Research Alliance, LLC under contract number DE-AC02-07CH11359 with the United States Department of Energy.
\end{acknowledgements}

\bibliography{bibliography}

\end{document}